\documentclass[twocolumn,aps,showpacs,longtable]{revtex4}

\usepackage{graphicx,latexsym}
\usepackage{epsfig}

\begin{document}
\title{Cloning a real $d$-dimensional quantum state on the edge 
of the no-signaling condition}
\author{Patrick Navez and Nicolas J. Cerf}
\affiliation{
Ecole Polytechnique, CP 165, Universit\'e Libre de Bruxelles, 1050
Brussels, Belgium}

\date{February 2003}
\begin{abstract}
We investigate a new class of quantum cloning machines 
that equally duplicate all real states in a Hilbert space 
of arbitrary dimension.
By using the no-signaling condition, namely that
cloning cannot make superluminal communication possible, 
we derive an upper bound on the fidelity of this class of 
quantum cloning machines.
Then, for each dimension $d$, we construct an optimal symmetric cloner whose 
fidelity saturates this bound.
Similar calculations can also be performed in order to recover 
the fidelity of the optimal universal cloner in $d$ dimensions. 
\end{abstract}

\maketitle
\section{Introduction}

The intimate connection between the impossibility of making 
a perfect quantum cloning machine (QCM) and the no-signaling 
condition, which prevents any superluminal communication, has
been realized since the seminal papers of Dieks\cite{Dieks} and
of Wootters and Zurek\cite{WZ}. More recently, Gisin has shown that
this connection can actually be exploited in order to recover the
fidelity 5/6 of the Buzek-Hillery universal QCM for qubits\cite{Gisin}.
Any cloning machine which would duplicate a qubit with a fidelity 
exceeding 5/6 would necessarily open a way to superluminal
communication. In short, the no-signaling condition is taken into account
by expressing that two statistical ensembles realizing the same
input density matrix (e.g. an equal mixture of $|0\rangle$ and
$|1\rangle$ or an equal mixture of $|0\rangle+|1\rangle$
and $|0\rangle-|1\rangle$) must result in indistinguishable output density
matrices for the clones. 
Since then,
this no-signaling constraint has also been
used to recover the fidelity of other classes of cloners, namely 
the asymmetric universal and phase-covariant qubit cloners\cite{Ghosh,Kunkri}.
\par

In this paper, we exploit this no-signaling condition 
in order to derive an upper bound on the fidelity 
of a new class of QCMs, which have not been considered in the
literature. We analyze symmetric QCMs which duplicate 
any $d$-dimensional {\em real} state with an equal fidelity.
These are the counterpart of the well-known universal QCMs
but within the realm of the so-called ``real'' quantum mechanics. 
We also find a constructive method to build QCMs that saturate this
upper bound, and therefore are optimal.
In particular, using this method for $d=2$, we find a cloner
unitarily equivalent to
the phase-covariant qubit cloner\cite{Bruss,Cerf-qutrit} 
which clones all states $a|0\rangle+b|1\rangle$ ($a$, $b$ being real
and satisfying $a^2+b^2=1$) with a fidelity $(1+1/\sqrt{2})/2$.
For an arbitrary dimension $d$, we use techniques from tensor calculus
in order to derive the no-signaling bound and the explicit cloner.
A specific application of this notion of real QCMs arises
in four dimensions, when these cloners are equivalent, 
up to a unitary transformation, to the universal cloners over 
the set of maximally entangled qubit pairs\cite{Lamoureux}.
\par

In general, the no-signaling requirement does not provide
a tight upper bound on the optimal cloning fidelity.
The linearity and trace preserving properties (which, combined, imply
the no signaling condition) need to be supplemented with 
the complete positivity
property in order to determine the best possible 
cloning transformation\cite{Bruss2}.
However, for the real QCMs of interest, it is sufficient
to combine the no-signaling requirement together with positivity 
(and trace preservation) in order to find a tight bound, 
just as in Gisin's original paper\cite{Gisin}. The impossibility
of signaling is crucial to derive this bound: 
would signaling be possible, an hypothetical perfect QCM 
providing two exact clones would then become permitted.
Furthermore, we show that a similar reasoning can also be applied 
in order to find the optimal fidelity of the universal QCM 
in $d$ dimensions\cite{Werner,BH,Cerf-qudit}. Thus, 
the requirement of no-signaling allows us 
to recover more simply and straightforwardly some standard results on cloning.
\par

A reason for which the no-signaling upper bound on the fidelity is saturated 
is that the set of states we are cloning is somehow ``large''. The set of 
$d$-dimensional real states is realized by means of the 
$SO(d)$ group representation, while the whole set of $d$-dimensional 
complex states is realized by means of the usual $SU(d)$ representation 
\cite{Hamermesh}. If we impose that the QCM acts equally on all the input
states defined by one of these representations, then the number of arbitrary 
independent parameters characterizing the cloning transformation 
is considerably reduced. 
For the real QCM, we will show that the density matrix 
can be reexpressed under the form of a covariant real tensor. 
This simplification allows us
to diagonalize the resulting density matrix and easily express 
no-signaling and positivity. Then, the initial optimization problem 
is turned into a simpler one involving only seven independent 
parameters, which can be solved analytically. 
\par

\section{No-signaling upper bound on the cloning fidelity}

The real input state to be cloned is defined in the computational
basis $\{|i\rangle\}$ as
\begin{eqnarray}
|\psi \rangle =
\sum_{i=0}^{d-1}
n_i |i \rangle
\end{eqnarray}
where the amplitudes $n_i$ are real and normalized 
as $\sum_{i=0}^{d-1} n_i^2=1$.
The two-clone output density matrix corresponding to this input 
state ${\bf n}=(n_0,\cdots, n_{d-1})$ is defined as
\begin{eqnarray}
\rho_{out}({\bf n})= \sum_{i,j,k,l=0}^{d-1}
r_{ij,kl}({\bf n}) \,
|i \rangle \langle k| \otimes |j \rangle \langle l|
\end{eqnarray}
We require that the QCM cloner act similarly on all real input 
states, that is
\begin{eqnarray}\label{cov}
\rho_{out}({\bf n'})=
U \otimes  U \, \rho_{out}({\bf n}) \, U^\dagger \otimes U^\dagger
\end{eqnarray}
where ${\bf n'}=(n_0',\cdots, n_{d-1}')$ with $n'_i=R_{ij} n_j$, and 
$U=\sum_{i,j}R_{ij}\,|i \rangle \langle j|$ is an arbitrary real rotation 
in the $d$-dimensional space satisfying 
$R_{ij}R_{kj}=\delta_{ij}$ (the summation symbol will be 
omitted from now on when dealing with tensors). 
This covariance property implies that $r_{ij,kl}({\bf n})$ is a tensor 
of rank four, i.e., it satisfies
\begin{eqnarray}\label{t4}
r_{ij,kl}({\bf n'})=
R_{ii'}R_{jj'}R_{kk'}R_{ll'} \, r_{i'j',k'l'}({\bf n}) 
\end{eqnarray}
Since we seek a symmetric cloner, the output 
density matrix must be invariant under the interchange of the two clones, 
i.e. under the permutations $i \leftrightarrow j$ and $k \leftrightarrow l$. 
The covariance and the permutation symmetry of the tensor impose
the following general form \cite{Hamermesh}:
\begin{eqnarray}\label{tensor}
r_{ij,kl}({\bf n})= 
\kappa_1 \delta_{ik}\delta_{jl} +
\kappa_2 \delta_{il}\delta_{jk}+
\kappa_3 \delta_{ij}\delta_{kl} +
\nonumber \\
\kappa_4 (n_i n_k \delta_{jl} + n_j n_l \delta_{ik}) +
\kappa_5 (n_i n_l \delta_{jk} + n_j n_k \delta_{il}) +
\nonumber \\
\kappa_6 (n_i n_j \delta_{kl} + n_k n_l \delta_{ij}) +
\kappa_7 n_i n_j n_k n_l
\end{eqnarray}
where the $\kappa_\alpha$ are seven independent real parameters. 
Note that if $\kappa_7=1$ and all other parameters vanish,
the two clones are perfect. The main result below  
is that the no-signaling condition imposes that $\kappa_7=0$,
so perfect cloning is precluded.

It is convenient, in what follows, to diagonalize this tensor,
Eq.~(\ref{tensor}),
and use its eigenvalues (along with a few other coefficients)
as independent parameters that characterize the tensor. The optimization
will then be made over these parameters. The diagonalization 
of Eq.~(\ref{tensor}) results in
\begin{eqnarray}\label{diag}
r_{ij,kl}({\bf n})=
\lambda_A  \sum_{\mu,\nu=1}^{d-1} V_{ij,\mu \nu}V^*_{kl,\mu\nu} 
+\lambda_B V_{ij}V^*_{kl}
\nonumber \\
+\lambda_C \sum_{\mu=1}^{d-1} V_{ij,\mu}V^*_{kl,\mu}
+\lambda_D \sum_{\mu=1}^{d-1} V'_{ij,\mu}V'^*_{kl,\mu}
\nonumber \\
+\lambda_E \sum_{
\begin{array}{c}
\mu_1, \dots, \mu_{d-3}=1 \\
\mu_1 \not= \dots \not= \mu_{d-3} \end{array} }
V_{ij,\mu_1 \dots \mu_{d-3}}V^*_{kl,\mu_1 \dots \mu_{d-3}} 
\end{eqnarray}
where the complete set of orthonormal eigenvectors 
are defined in Table I. 

\begin{table*}[h]
\begin{tabular}{ccc}
\hline
Eigenvalue & Eigenvectors & Degeneracy \\
\hline
$\lambda_A$ & 
$V_{ij,\mu \nu}=\displaystyle{\cos\alpha\sin(2\phi)\left[\frac{1}{\sqrt{d-1}}
n_i n_j+
\frac{\cot\phi}{d-1}\left(n_i n_j - \delta_{ij} \right)\right]
\delta_{\mu\nu}+ \frac{e^{i\alpha}}{2}
\left(m_k^\mu m_l^\nu + m_k^\nu m_l^\mu\right)}$ 
& $\displaystyle{\frac{d(d-1)}{2}}$ \\
$\lambda_B$ &
$V_{ij}=\displaystyle{\cos\phi \, n_i n_j- \frac{\sin \phi}{\sqrt{d-1}}
\left(n_i n_j - \delta_{ij} \right)}$& $1$ \\
$\lambda_C$ &
$V_{ij,\mu}=\cos \theta n_i m_j^\mu + \sin \theta m_i^\mu n_j$& $d-1$\\
$\lambda_D$ &
$V'_{ij,\mu}=-\sin \theta  n_i m_j^\mu +\cos \theta m_i^\mu n_j$& $d-1$ \\
$\lambda_E$ &
$V_{ij,\mu_1 \dots \mu_{d-3}} = \displaystyle{
\sum_{ \mu_{d-2}, \mu_{d-1}=1 }^{d-1}
\frac{1}{\sqrt{2}} \epsilon_{\mu_1, \dots, \mu_{d-1}}
m_i^{\mu_{d-1}} m_j^{\mu_{d-2}}}
$& 
$\displaystyle{\frac{(d-1)(d-2)}{2}}$\\
\hline
Total && $d^2$ \\
\hline
\end{tabular}
\caption{Eigenvector decomposition of the rank-four 
$d^2\times d^2$ tensor $r_{ij,kl}$
characterizing the two-clone density matrix of the real QCM in $d$
dimensions when the input state is $n_i$.}
\end{table*}

Note that all the eigenvectors are normalized to unity except for the
off-diagonal eigenvectors of the symmetric subset $V_{ij,\mu \nu}$
which are normalized to 1/2, i.e.,
$\sum_{i,j} V_{ij,\mu \nu}V^*_{ij,\mu'\nu'}= (\delta_{\mu\mu'}\delta_{\nu\nu'}
+\delta_{\mu\nu'} \delta_{\nu\mu'})/2 $.
Here, the coefficients $m_i^\mu$ denote an arbitrary set of basis vectors 
($1 \leq \mu \leq d-1$) of the subspace orthogonal to ${\bf n}$. 
The notation $\epsilon_{\mu_1, \dots, \mu_{d-1}}$ stands for the unit 
antisymmetric tensor of rank $d-1$, which is equal to $1$ 
if $(\mu_1, \dots, \mu_{d-1})$ 
is an even permutation of $(1,\dots, d-1)$, to $-1$ if 
 $(\mu_1, \dots, \mu_{d-1})$ is an odd permutation of 
$(1,\dots, d-1)$, and to $0$ if any index is repeated. 
The permutation symmetry between the two clones imposes that 
$(\lambda_C-\lambda_D)\cos(2\theta)=0$, so that either 
$\lambda_C=\lambda_D$ or $\cos \theta = \pm 1 /\sqrt{2}$.  
This constraint reduces to seven the number of independent parameters 
among the eight parameters $\lambda_I$ ($I=A,B,C,D,E$), $\alpha$, $\gamma$, 
and $\theta$. A straightforward identification between
expressions (\ref{tensor}) and (\ref{diag}) allows us to unambiguously express
the seven independent parameters $\kappa_\alpha$ in terms of the new ones.

Let us now consider the density matrix of each of the two clones
and their fidelity with respect to the input state.
The two clones are in the same mixture due to permutation symmetry, and the 
covariance imposes that the density matrix is given by a rank-two tensor of the form
\begin{eqnarray}
&&{\rm Tr}_1 \rho_{out}({\bf n})= {\rm Tr}_2 \rho_{out}({\bf n})
\nonumber\\
&=&\sum_{i,j=0}^{d-1}\frac{1}{d-1}\left[
(dF-1)n_i n_j +(1-F) \delta_{ij} \right]
|i\rangle \langle j| 
\nonumber \\
&=& F |\psi \rangle \langle \psi |+
\frac{1-F}{d-1}\left(\openone-|\psi \rangle \langle \psi |\right)
\end{eqnarray}
where $F$ is the fidelity
\begin{eqnarray}
F={\rm Tr}_1\left(|\psi \rangle \langle \psi | \rho_{out}({\bf n})\right)
= n_i n_k r_{ij,kj}({\bf n})
\end{eqnarray}
Using Eq.~(\ref{diag}), we can express the fidelity in terms of 
the eigenvalues and  eigenvector parameters,
\begin{eqnarray}\label{F}
F=\lambda_A \cos^2 \alpha \sin^2(2 \phi)+ 
\lambda_B \cos^2 \phi  
\nonumber \\
+ (\lambda_C \cos^2 \theta +
\lambda_D \sin^2 \theta)(d-1)   \, .
\end{eqnarray}
When maximizing $F$, we will have to take into account 
the three following constraints:

{\it i) Positivity:} $\rho_{out}\ge 0$ \\
This gives $\lambda_I \geq 0$ with 
$I=A,B,C,D,E$.

{\it ii) Trace preservation}: ${\rm Tr}(\rho_{out})=1$\\
\begin{eqnarray}\label{ut}
\frac{d(d-1)}{2}\lambda_A + \lambda_B + 
(d-1)(\lambda_C+\lambda_D) 
\nonumber \\
+ \frac{(d-1)(d-2)}{2}\lambda_E=1
\end{eqnarray}

{\it iii) No-signaling condition}: \\
This requires that the uniform mixture of any two basis sets ${\bf n}^\mu$ 
and ${{\bf n}^\mu}'$ (which thus both realize the same input density matrix,
namely the identity) must result in two indistinguishable output
density matrices. Thus, 
\begin{eqnarray}\label{ns}
\sum_{\mu=0}^{d-1} \rho_{out}({\bf n}^\mu)=
\sum_{\mu=0}^{d-1} \rho_{out}({\bf n'}^\mu)
\end{eqnarray}
Using Eq.~(\ref{tensor}) and the 
completion relation $\sum_{\mu=0}^{d-1}n_i^\mu n_j^\mu
=\sum_{\mu=0}^{d-1}{n_i^\mu}' {n_j^\mu}' = \delta_{ij}$, 
the only way of satisfying Eq.~(\ref{ns}) is to forbid quartic term in 
(\ref{tensor}), i.e. to 
impose $\kappa_7=0$. As mentioned earlier, this means that the
``perfect cloning'' term in Eq.~(\ref{tensor}) is forbidden.
If this no-signaling condition is not obeyed, then by maximizing $F$ we 
obtain a perfect cloner described in terms of the only 
eigenvector $V^{ij}$ setting $\phi=0$. Thus, according to intuition, 
we observe that this no-signaling condition 
is needed in order to exclude perfect cloning. 
In terms of the eigenvalues and eigenvector parameters, 
this no-signaling condition becomes
\begin{eqnarray}\label{nsc}
\lambda_A t_A
+ \lambda_B t_B 
=\lambda_C \left(1+ \sin 2\theta\right) +
\lambda_D \left(1 - \sin 2\theta\right)  \, ,
\end{eqnarray}
where we have defined the positive coefficients
\begin{eqnarray}
t_A=\cos^2 \alpha \left(\frac{d-2}{d-1}\sin^2(2\phi)
+\frac{\sin(4\phi)}{\sqrt{d-1}}\right)
+1 \geq 0
\end{eqnarray}
\begin{eqnarray}
t_B=\left(\cos \phi - \frac{\sin \phi}{\sqrt{d-1}}\right)^2
\geq 0
\end{eqnarray}

Now, the constrained optimization problem can be solved analytically
in order to upper bound the cloning fidelity.
First, we observe that when $\lambda_E \not=0$, we can always 
increase the fidelity by substituting $\lambda_I$ ($I=A,B,C,D$) 
with $\lambda_I /(1 - (d-1)(d-2)\lambda_E/2)$ and $\lambda_E$ 
with $0$. This substitution increases the fidelity while keeping 
the constraints satisfied. Therefore, the requirement 
$\lambda_E=0$ always gives an optimal fidelity. 
Second, remember that the permutation symmetry imposes 
either (a) $\cos \theta = \pm 1 /\sqrt{2}$ or (b) $\lambda_C=\lambda_D$.
We will consider these two possibilities.

{\it Case (a).} Let us examine first the case $\cos \theta = 1/\sqrt{2}$.
We eliminate the 
variable $\lambda_C$ between Eqs.(\ref{ut}) and (\ref{nsc}), resulting in
\begin{eqnarray}\label{ut2}
\frac{d-1}{2}\left(d + t_A \right)\lambda_A +
\left(1+\frac{d-1}{2}t_B\right)\lambda_B +(d-1)\lambda_D=1
\nonumber \\
\end{eqnarray}
Similarly, combining Eqs.~(\ref{F}) and (\ref{nsc}) gives
\begin{eqnarray}\label{F2}
F=\left(\cos^2\alpha \sin^2(2\phi)+\frac{d-1}{4}t_A\right)\lambda_A
\nonumber \\
+\left(\cos^2\phi +\frac{d-1}{4}t_B\right)\lambda_B +
\frac{d-1}{2}\lambda_D
\end{eqnarray}
The coefficients in front of the eigenvalues 
$\lambda_A$, $\lambda_B$, and $\lambda_D$
are all semi-positive in Eqs.(\ref{ut2}) and (\ref{F2}), so that
only one of these eigenvalues is non-zero in the optimum.
For each non-zero eigenvalue, Eqs.(\ref{ut2}) and
(\ref{F2}) give a value for the fidelity, and
the maximum fidelity is simply chosen as the best of these 
three possibilities. We find that the fidelity is upper bounded by
\begin{eqnarray}\label{F4}
\max \left\{
\frac{\cos^2\alpha \sin^2(2\phi)+\frac{d-1}{4}t_A}
{\frac{d-1}{2}\left(d + t_A \right)},
\frac{\cos^2\phi +\frac{d-1}{4}t_B}
{1 +\frac{d-1}{2}t_B},\frac{1}{2} 
\right\}
\nonumber \\
\end{eqnarray}
The first term in the maximum, Eq.(\ref{F4}), 
must be greater than $1/2$ to be of interest.
This condition is fulfilled only if 
$\cos^2\alpha \sin^2(2\phi)\geq d(d-1)/4$ and this can be the case 
only when $d=2$. But, for $d=2$, we notice that the first term 
is maximized by choosing $\cos(\alpha)=1$, since
the optimum always lies within the range $\pi \leq 4\phi \leq 3\pi/2$. 
Moreover, if we substitute $2\phi$ with $- \phi$, we recover the 
second term of (\ref{F4}). Thus, optimizing the first term 
for $d=2$ amounts to optimizing the second term. As a 
consequence, we are left with maximizing the second term 
of (\ref{F4}) for any dimension, which only depends on $\phi$. 
The maximum is found for 
\begin{eqnarray}
\tan \phi=
\frac{d+4 - \sqrt{d^2+4d +20}}{2\sqrt{d-1}}
\end{eqnarray}
Consequently, the cloning fidelity of the real QCM in $d$ dimensions
cannot exceed the following upper bound 
\begin{eqnarray}\label{Ffinal}
F \leq F_{max} =  \frac{1}{2}+
\frac{\sqrt{d^2+4d +20} -d +2}{4(d+2)}
\end{eqnarray}
in order to make signaling via cloning impossible.
This is the main result of this Section.

{\it Case (b).} In order to be complete, let us consider the second case 
$\lambda_C = \lambda_D$ and show that the upper bound cannot be improved.
Similarly to the first case, we 
eliminate the variable $\lambda_C$ 
from Eqs.(\ref{ut}) and (\ref{nsc}), and obtain equations 
similar to Eqs. (\ref{ut2}) and (\ref{F2}), namely 
\begin{eqnarray}\label{ut3}
\frac{d-1}{2}\left(d + 2 t_A \right)\lambda_A +
\left(1+(d-1)t_B\right)\lambda_B=1
\end{eqnarray}
and
\begin{eqnarray}\label{F5}
F=\left(\cos^2\alpha \sin^2(2\phi)+\frac{d-1}{2}t_A\right)\lambda_A
\nonumber \\
+\left(\cos^2\phi +\frac{d-1}{2}t_B\right)\lambda_B 
\end{eqnarray}
We then obtain an upper bound on $F$ given by
\begin{eqnarray}\label{F6}
\max \left\{
\frac{\cos^2\alpha \sin^2(2\phi)+\frac{d-1}{2}t_A}
{\frac{d-1}{2}\left(d + 2t_A \right)},
\frac{\cos^2\phi +\frac{d-1}{2}t_B}
{1 +(d-1)t_B}
\right\}
\nonumber \\
\end{eqnarray}
From (\ref{F6}), we note that for the fidelity 
to be greater than $1/2$, then either 
$\cos^2\alpha \sin^2(2\phi) > 1/2$ or $\cos^2\phi > 1/2$.
But if one of these conditions is satisfied, then each term 
in Eq.~(\ref{F6}) is lower than the corresponding one in 
Eq.~(\ref{F4}). Therefore, we conclude that the no-signaling upper bound 
is indeed given by (\ref{Ffinal}).

\section{Real QCM saturating the no-signaling bound}


We will now explicitly construct a real QCM
and observe that it saturates the no-signaling upper 
bound Eq.~(\ref{Ffinal}). Hence, we will have found an optimal real 
QCM in $d$ dimensions.
We will follow here the constructive method described in
Ref.~\cite{Cerf-qudit}, which consists in considering 
the cloning of an input system that is maximally entangled 
with a reference system denoted as $R$, i.e.,
$\sum_{i=1}^d |i\rangle |i\rangle$. In this case, the joint state 
$|\Psi\rangle_{R,1,2,A}$ of the reference, 
the two output clones, and the ancilla completely
characterizes the cloning transformation. (The reference and
ancilla systems are assumed to belong to a space of dimension $d$,
just as the input and the two clones.)
We consider here the most general state 
\begin{eqnarray}
|\Psi\rangle_{R,1,2,A}= \sum_{i,j,k,l=0}^{d-1}
u_{ijkl}|l \rangle_I |i\rangle_1 |j \rangle_2 |k \rangle_A
\end{eqnarray}
where the indexes 1, 2, $R$ and $A$ refer respectively to the two
clones, the reference, and the ancilla.
Note that this state does not depend on ${\bf n}$, but it can be easily used
in order to define the cloning transformation applied on state $|\psi\rangle$:
projecting the reference system of $|\Psi\rangle_{I,1,2,A}$ 
onto $|\psi\rangle$\footnote{In general, one should project the reference onto 
the complex conjugate of the input state $| \psi^*\rangle$, but
this is irrelevant here since we deal with real QCMs.}
amounts to defining the cloning transformation as:
\begin{eqnarray}\label{ct}
|\psi \rangle \rightarrow \sum_{i,j,k,l=0}^{d-1}
u_{ijkl}\; n_l \; |i\rangle_1 |j \rangle_2 |k \rangle_A
\end{eqnarray}
We require that this state obeys the following covariance principle
\begin{eqnarray}
|\Psi\rangle_{R,1,2,A} = 
U^* \otimes U \otimes U \otimes U^* \; |\Psi\rangle_{R,1,2,A}
\end{eqnarray}
for all real unitary rotations $U$ as those used in Eq.~(\ref{cov}).
This strong requirement allows to recover the property (\ref{cov}), 
while the converse is not necessarily true. The condition (\ref{ct}) 
physically means that applying a rotation $U$ on the input (or
rotating the reference by $U^*$) is equivalent to rotating the two
clones by $U$ and the ancilla by $U^*$.
This covariance principle implies that $u_{ijkl}$ is a tensor of rank
four, that is, it satisfies
\begin{eqnarray}\label{u4}
u_{ijkl}=
R_{ii'}R_{jj'}R_{kk'}R_{ll'}u_{i'j'k'l'}
\end{eqnarray}
The most general tensor obeying Eq.~(\ref{u4}) can be
written as \cite{Hamermesh}:
\begin{eqnarray}
u_{ijkl}=A  \delta_{il} \delta_{jk}+ B \delta_{jl} \delta_{ik} +
C \delta_{kl} \delta_{ij}
\end{eqnarray}
The symmetry permutation between the two clones imposes 
that $A=B$. The fidelity $F$ can be obtained from Eq.~(\ref{ct})
by tracing over one of the clones and the ancilla,
resulting in
\begin{eqnarray}
F = (d+3)|A|^2 + |C|^2 +2(A C^* + C A^*)
\end{eqnarray}
This expression has to be maximized under the normalization constraint
\begin{eqnarray}
2(d+1)|A|^2+d|C|^2 +2(A C^* + C A^*)=1
\end{eqnarray}
It can be checked that this maximization procedure exactly gives 
the right-hand side of Eq.~(\ref{Ffinal}), so that the cloner we have
constructed saturates the no-signaling bound. 
The corresponding optimal coefficients are given by
\begin{eqnarray}
A&=&\left[\frac{d\sqrt{d^2+4d +20}+2d+d^2-8}
{4(d-1)(d+2)\sqrt{d^2+4d +20}}\right]^{1/2}
\\
C&=&\frac{\sqrt{d^2+4d +20}-d-2}{4}A
\end{eqnarray}

\section{Case of the $d$-dimensional universal cloner}

Consider now a universal cloner, that is a QCM such that
any pure state $|\psi\rangle= \sum_{i=0}^{d-1} c_i|i\rangle$ 
(with $c_i$ being complex amplitudes)
is cloned with the same fidelity. To obey the 
covariance properties (\ref{cov}) for {\em any} unitary transformation 
$U$, the output density matrix must have the more restricted form:
\begin{eqnarray}\label{tensorc}
r_{ij,kl}(c_i)=
\kappa_1 \delta_{ik}\delta_{jl} +
\kappa_2 \delta_{il}\delta_{jk}+
\kappa_4 (c_i c^*_k \delta_{jl} + c_j c^*_l \delta_{ik}) 
\nonumber \\
+ \kappa_5 (c_i c^*_l \delta_{jk} + c_j c^*_k \delta_{il}) +
\kappa_7 c_i c_j c^*_k c^*_l
\end{eqnarray}
In comparison with (\ref{tensor}), the covariance condition imposes that 
$\kappa_3=\kappa_6=0$ and, consequently, that $\phi=0$. 
As a result, the second term in (\ref{F4}) 
gives a smaller upper bound so that we find
\begin{eqnarray}\label{Ffinalu}
F \leq
\frac{1}{2}+
\frac{1}{d+1}
\end{eqnarray}
The universal $d$-dimensional cloner saturating this bound
has been discussed in Refs.~\cite{Werner,BH,Cerf-qudit},
so we see that the no-signaling condition again gives a tight bound.
We can recover this cloner by following Section III. The
covariance condition implies that $C=0$ and,
as a consequence, $A=1/\sqrt{2(d+1)}$.

\section{Discussion and conclusion}

We have found a new class of QCMs which duplicate any $d$-dimensional 
real state with an equal fidelity
\begin{eqnarray}
F = \frac{1}{2}+ \frac{\sqrt{d^2+4d +20} -d +2}{4(d+2)}  \; .
\end{eqnarray}
Furthermore, for these universal cloners over real states in $d$ dimensions, 
we have demonstrated that the no-signaling requirement provides a sufficient constraint to unambiguously determine the optimal performance of the cloners.
Without this no-signaling constraint, we would obtain a perfect cloner forbidden by quantum theory. Hence, we have found the optimal real QCMs.
\par

In the special case of $d=2$, we recover the phase-covariant 
qubit cloner of fidelity 
\begin{eqnarray}
F_{d=2}={1+1/\sqrt{2}\over 2}\simeq 0.854
\end{eqnarray}
as derived in \cite{Bruss}
(see also the Appendix of \cite{Cerf-qutrit}). 
For qutrits ($d=3$), we get a cloner of fidelity 
\begin{eqnarray}
F_{d=3}={9+\sqrt{41}\over 20}\simeq 0.770
\end{eqnarray}
This result is distinct from the fidelity of the three known QCMs 
for qutrits\cite{Cerf-qutrit}: $F=3/4$ for the universal qutrit
cloner, $F=(5+\sqrt{17})/12\simeq 0.760$ for the two-phase covariant
qutrit cloner, and $F=1/2+1/\sqrt{12}\simeq 0.789$ for the qutrit cloner
of two mutually unbiased bases. This suggests that these real QCMs
form a genuinely new class of QCMs. 
Note, finally, that when the dimension $d$ tends to infinity, 
the cloning fidelity $F$ tends to $1/2 +O(1/d)$.
In Fig. 1, we have plotted, for comparison, the fidelity 
as a function of the dimension $d$ for the universal cloner, 
the real cloner derived here, and the optimal cloner 
of two mutually unbiased bases obtained in Ref.~\cite{Boure}.
As expected, we observe that the real QCM has a higher fidelity than the
universal QCM since it clones the restricted class of real states. 
However, the real QCM performs less well 
than the cloner of two mutually unbiased bases (except when $d=2$
where they coincide).

\begin{figure}
\scalebox{0.6}{\includegraphics{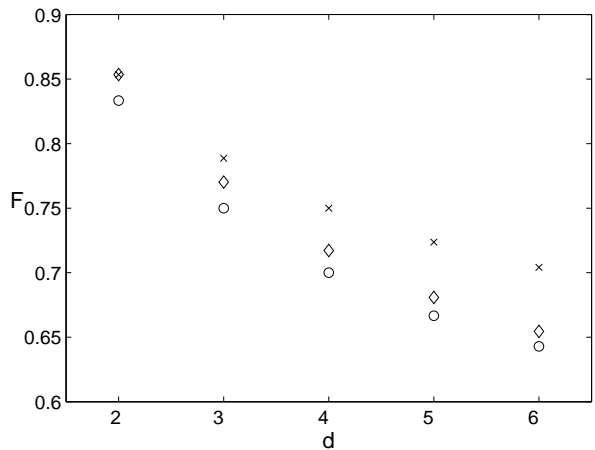}}
\caption{Fidelity $F$ as a function of the dimension $d$ 
for the universal cloner ($o$) \cite{Werner,BH,Cerf-qudit}, 
the real cloner derived in the present paper ($\Diamond$),
and the cloner of two mutually unbiased bases ($\times$) \cite{Boure}.}
\end{figure}

An interesting issue of this work is the potential generalization 
of this method exploiting the no-signaling constraint
to any kind of cloners. It is likely, however, that for 
a more restrictive set of states to be cloned equally,  
the no-signaling constraint may only give a non-tight upper 
bound for the fidelity. A typical example may be cloning the set 
of only two mutually unbiased bases: this smaller set might impose 
a weaker covariance constraint to the cloning transformation,
so the maximum fidelity consistent with no-signaling might correspond to a cloner that is not allowed by quantum mechanics. This will be further investigated.

\bigskip
\centerline{\bf ACKNOWLEDGMENTS}

We thank Sofyan Iblisdir and Louis-Philippe Lamoureux  
for many helpful discussions.
We also acknowledge funding from the European Union under the
project RESQ (IST-FET programme) and from the Communaut\'e
Fran\c caise de Belgique under the ``Action de Recherche Concert\'ee''
nr. 00/05-251.

\end{document}